\documentclass[superscriptaddress,twocolumn,showpacs,prl]{revtex4}
\usepackage{graphicx}

\begin{document}

\title{
Size dependence of the minimum excitation gap in the Quantum Adiabatic
Algorithm}

\author{A.~P.~Young}
\email{peter@physics.ucsc.edu}
\affiliation{Department of Physics, University of California,
Santa Cruz, California 95064}

\author{S.~Knysh}
\affiliation{ELORET Corporation, NASA Ames Research Center, MS 229,
Moffett Field, CA 94035-1000}
\email{sergey.i.knysh@nasa.gov}

\author{V.~N.~Smelyanskiy}
\affiliation{NASA Ames Research Center, MS 269-3, Moffett Field, CA
94035-1000}
\email{Vadim.N.Smelyanskiy@nasa.gov}

\date{\today}

\begin{abstract}
We study the typical (median) value of the
minimum gap in the quantum version of the Exact Cover
problem using Quantum Monte Carlo simulations, in order to
understand the complexity of the quantum adiabatic algorithm (QAA) for much
larger sizes than before.  For a range of sizes, $N \le
128$, where the classical Davis-Putnam algorithm shows exponential median
complexity, the QAA shows polynomial median complexity.
The bottleneck of the algorithm is an isolated avoided crossing point of
a Landau-Zener type (collision between the two lowest energy levels
only).
\end{abstract}
 \maketitle

There is considerable interest in finding optimization problems which
could be solved much more efficiently by an eventual quantum computer
than by a classical computer. An important class of classically
intractable problems is the NP-hard category \cite{Garey:97}.  Many
optimization problems of current interest have parameters which are
random and so each problem corresponds to a large number (possibly
infinite) of ``instances''. The term NP-hard actually refers to the
behavior of the computationally \textit{hardest} instance, but, from a
practical point of view, it is also of great interest to know how the
time to solve a \textit{typical} instance~\cite{dubois:01,hogg:96}, the
typical complexity, scales with problem size. Numerical studies of
NP-hard problems show that this scaling is exponential in a broad class
of problem parameters \cite{hogg:96,dubois:01}. It would be a very
important breakthrough to show that a quantum computer can solve the
same class of problem instances of an NP-hard problem in less then
exponential time.

In this paper we study the typical complexity as a function of system
size for a particular quantum algorithm, the quantum adiabatic algorithm
(QAA) proposed by Farhi et al.~\cite{farhi:01b}. The idea is that one
adds to a \lq\lq problem" Hamiltonian, $\mathcal{H}_{\rm P}$, whose
ground state represents a solution of a classical optimization problem
a non-commuting \lq\lq driver" Hamiltonian, $\mathcal{H}_{\rm D}$, so
the total Hamiltonian is
\begin{equation}
\mathcal{H}(\lambda) = (1-\lambda) \mathcal{H}_{\rm D} + \lambda \mathcal{H}_{\rm P}, \label{qu_ham}
\end{equation}
where $\lambda\equiv\lambda(t)$ is a \textit{time dependent} control
parameter. For $\mathcal{H}_{\rm P}$ we
are interested in binary optimization problems expressed
in terms of classical Ising spins taking values $\pm 1$, or equivalently
in terms of the $z$-components of
the Pauli matrices for each spin, $\hat\sigma^z_i$. The driver
Hamiltonian is then simply $\mathcal{H}_{\rm D} = -\sum_{i=1}^N
\hat\sigma^x_i$ where $\hat\sigma^x_i$ is the $x$-component Pauli matrix.

The control parameter $\lambda(t)$ is 0 at $t=0$, so
$\mathcal{H}$=$\mathcal{H}_{\rm D}$, which has a trivial ground state in
which all $2^N$ basis states (in the $\hat\sigma^z$ basis) have equal
amplitude. It then increases with $t$, reaching 1 at $t={\cal T}$
(${\cal T}$ is the runtime or complexity of the algorithm), at which
point $\mathcal{H}$=$\mathcal{H}_{\rm P}$. If the time evolution of
$\lambda(t)$ is sufficiently slow, the process will be adiabatic. Hence,
starting the system in the ground state of $\mathcal{H}_{\rm D}$ (all
spins aligned along $x$), the system will end up in the classical ground
state, which is what we want, with only small probability of failure.
An upper bound for the complexity of the QAA can be
given~\cite{wannier:65,farhi:02}, in terms of the eigenstates and
eigenvalues of the Hamiltonian, ${\cal H}\Phi_n=E_n\Phi_m$,
\begin{equation}
{\cal T}\gg  \hbar |\max_{\lambda}V_{10}(\lambda)| / \left(\Delta E_{\rm min}\right)^2, \label{bound}
\end{equation}
where $\Delta E_{\rm min}$ corresponds to the minimum of the first
excitation gap $\Delta E_{\rm min}=\min_{\lambda}\Delta E(\lambda)$ with
$\Delta E=E_1-E_0$, and  $V_{n0}(\lambda)=\langle
\Psi_0|d\mathcal{H}/d\lambda|\Psi_n\rangle$. Typically, matrix elements
of $\mathcal{H}$  scale as a low polynomial of a number of spins $N$
and the question of whether the complexity ${\cal T}$ depends
polynomially or exponentially with $N$ depends on how the minimum gap
$\Delta E_{\rm min}$ scales with $N$. The size dependence of the minimum
gap will therefore be the central focus of this paper.

It is difficult to
study the typical complexity of the QAA \textit{analytically}
since
$\lambda^*$, the value of $\lambda$ at the minimum of the gap $\Delta
E(\lambda)$, is different for each instance with fluctuations being
${\cal O}(N^{-1/2})$, so the ensemble averaging over random instances
can only be performed {\it after} $\lambda^*$ has been found for each
case.
In the original work of Farhi et al.~\cite{farhi:01b}, the complexity of
the adiabatic algorithm was studied \textit{numerically}
by direct integration in
time of the system with Hamiltonian ${\cal H}$. Since the size of the
Hilbert space increases exponentially (it is of order $2^N$) they were
limited to very small sizes, $N \lesssim 20$. Subsequently
Hogg~\cite{hogg:03} considered sizes up to $N = 24$.
These
early papers~\cite{farhi:01b,hogg:03} found that the complexity of the
algorithm scales as a roughly as $N^2$. However, this power law complexity
may be an artifact of the very small sizes studied, so it is of great
interest to determine whether the complexity continues to be polynomial
for much larger sizes or whether a ``crossover'' to exponential
complexity is seen.  To investigate this question, it is not possible to
include all terms in the Hilbert space (as was done in the early work)
since this becomes much too large.
Here we use
Quantum Monte Carlo (QMC) simulations,
with which we can study much larger sizes
because only a \textit{sampling} of the states is performed.

There have also been  QMC simulations, see e.g.~Ref.~\cite{santoro:06}
for a discussion, in which $t$ in Eq.~(\ref{qu_ham}) is the number of
Monte Carlo sweeps, and one estimates how the final excess energy
(i.e.~the energy above the ground state) varies with the \textit{total}
number of sweeps $\mathcal{T}$.  However, this is a ``fake'' dynamics,
which is not necessarily representative~\cite{santoro:06} of the real
time unitary evolution  guided by the Schr\"odinger equation.  Therefore
the computational complexity of such a procedure does not necessarily
correspond to that of the quantum adiabatic algorithm~\cite{farhi:01b}.

\begin{figure}
\begin{center}
\includegraphics[width=\columnwidth]{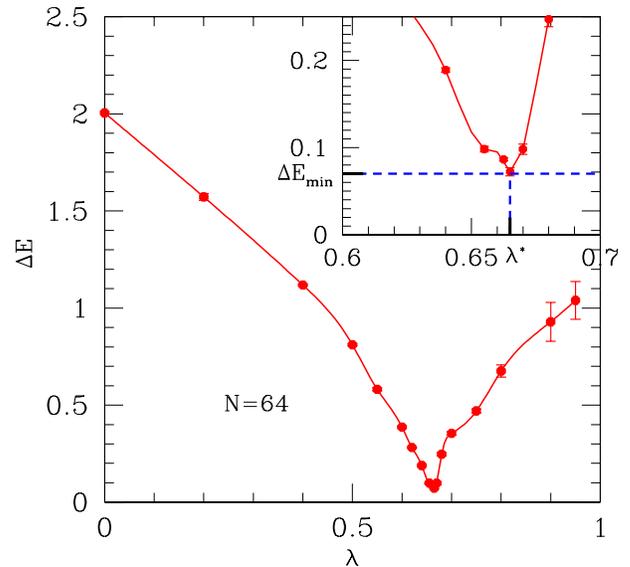}
\caption{(Color online) QMC results for the gap between the ground state and the first
excited state as a function of the control parameter $\lambda$ for one instance with $N=64$. The region
around the minimum value of the gap, $\Delta E_{\rm min}$, which occurs at $\lambda = \lambda^*$, is blown
up in the inset. }
\label{fig:DE_1}
\end{center}
\end{figure}

To make a comparison with the earlier work we study (essentially) the
same model of $\mathcal{H}_{\rm P}$ used by Farhi et
al.~\cite{farhi:01b}. It corresponds to an Exact Cover problem, which is a
particular version of a Constraint Satisfaction, a commonly studied
problem in the NP-hard category. In Exact Cover there are $N$ Ising
spins  and $M$ ``clauses'' each of which involves three spins (chosen at
random). The energy of a clause is zero if one spin is $-1$ and the other
two are $1$, otherwise the energy is 1. Thus $\mathcal{H}_{\rm P}$ equals
\begin{eqnarray}
\mathcal{H}_{\rm P}  & = & {1 \over 8} \, \sum_{\alpha=1}^M  \Big( 5 - \hat\sigma^z_{\alpha_1} -
\hat\sigma^z_{\alpha_2} - \hat\sigma^z_{\alpha_3} + \hat\sigma^z_{\alpha_1}\, \hat\sigma^z_{\alpha_2}
\nonumber \\
& + & \hat\sigma^z_{\alpha_2}\, \hat\sigma^z_{\alpha_3} +\hat\sigma^z_{\alpha_3}\, \hat\sigma^z_{\alpha_1} + 3\,
\hat\sigma^z_{\alpha_1}\, \hat\sigma^z_{\alpha_2}\, \hat\sigma^z_{\alpha_3} \Big) \, , \label{hclass}
\end{eqnarray}
where $\alpha_1, \alpha_2$ and $\alpha_3$ are the three spins in clause
$\alpha$ and the $\{\hat\sigma^{z}_i\}_{i=0}^{i=N}$ are Pauli matrices.
In the absence of the driver Hamiltonian, the Pauli matrices can be
replaced by classical Ising spins taking values $\pm 1$.
An
instance has a ``satisfying assignment'' if there is at least one choice
for the spins where the total energy is zero. As the ratio $M/N$ is
increased, there is a phase transition where the number of satisfying
assignments goes to zero.  The version used by Farhi et al.~considers
only instances with a \textit{unique} satisfying assignment (USA), i.e.
there is only \textit{one} state with energy 0. This has the advantage
that the gap $\Delta E(\lambda)$ is greater than zero in both limiting cases,
${\cal H}=\mathcal{H}_{\rm D}$ and ${\cal H}=\mathcal{H}_{\rm P}$, but
will have a minimum at an intermediate value $\lambda = \lambda^*$, see
Fig.~\ref{fig:DE_1}. The aim is to determine the size $N$ dependence of
the typical value of $\Delta E_{\rm min}$, averaged over many instances.

\begin{figure}
\begin{center}
\includegraphics[width=\columnwidth]{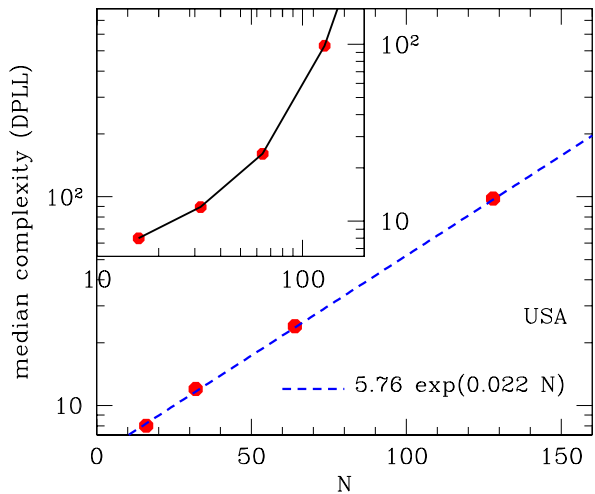}
\caption{(Color online) A log-linear plot of the median complexity of
the exact cover problem using the (classical) DPLL algorithm as a
function of $N$. The straight line fit works well demonstrating that the
\textit{complexity increases exponentially} with $N$ even for quite
modest sizes.  This figure is for samples with a USA but the data for
all samples (with the same number of clauses $M$) is very similar. The
inset plots the same data on a log-log scale. The pronounced curvature
shows that the data can \textit{not be fitted to a power law}. }
\label{fig:DPLL}
\end{center}
\end{figure}

We generate instances with a USA
as follows. For each size $N$,
we take $M$ clauses and prune off (i) isolated sites, and (ii) clauses
(think of them as triangles) which are only connected to other clauses
at one corner, since these give a trivial degeneracy without changing
the complexity. This leaves $N'$ sites and $M'$ clauses. Using the
standard Davis-Putnam-Logemann-Loveland (DPLL) ~\cite{davis:62}
algorithm we then see if the remaining $N'$ sites with $M'$ clauses have
a USA. For each $N$, we choose $M$ to maximize the probability of
finding a USA. Although the probability of finding a USA decreases
exponentially with $N$, we have easily been able to find instances for
$N$ up to 256 and the values of $M$ are shown in Table \ref{tab:M}. For
the sizes which we will study by QMC ($N \le 128$) the DPLL algorithm
clearly shows exponential complexity, see Fig.~\ref{fig:DPLL}.

\begin{table}
\begin{center}
\begin{tabular}{|r|r|r|r|r|r|r|r|}
\hline\hline
N        & 16     & 32     & 64     & 128    & 192    & 256 \\
\hline
M        & 12     & 23     & 44     & 86     & 126    & 166 \\
\hline
$\alpha$ & 0.7500 & 0.7188 & 0.6875 & 0.6719 & 0.6563 & 0.6484 \\
\hline\hline
\end{tabular}
\end{center}
\caption{For sizes $N$ up to 256 we show values of the number of clauses
$M$ for which the probability of a unique satisfying assignment (USA),
constructed as described in the text, is maximized. The ratio $M/N$ is
denoted by $\alpha$, and is expected to approach the value at the
quantum phase transition $\alpha_c \simeq 0.625$~\cite{knysh:04} for $N
\to \infty$. For the QMC simulations we only used the sizes up to $N =
128$.}
\vspace{-0.5cm}
\label{tab:M}
\end{table}

For each instance, we use QMC to simulate the quantum system in
Eqs.~(\ref{qu_ham}) and (\ref{hclass}) with  $N'$ spins and $M'$
clauses.
We simulate an
effective classical model with Ising spins $\sigma^z_i(\tau)=\pm1$ in
which
$\tau$ ($0 \le \tau < \beta \equiv T^{-1}$) is
imaginary time.
In practice, imaginary time
is discretized into $L_\tau$ ``time slices'' each representing $\Delta
\tau = \beta/L_\tau$ of imaginary time.  
For, a different model, the 1-d Ising chain in a transverse field
we have verified numerically~\cite{knysh:08}
that the scaling behavior of the energy gap~\cite{fisher:95} is the same
for $\Delta\tau\to 0$ as for finite $\Delta\tau$, and hence it is plausible
that a discrete $\Delta\tau$ will work here too.

\begin{figure}
\begin{center}
\includegraphics[width=\columnwidth]{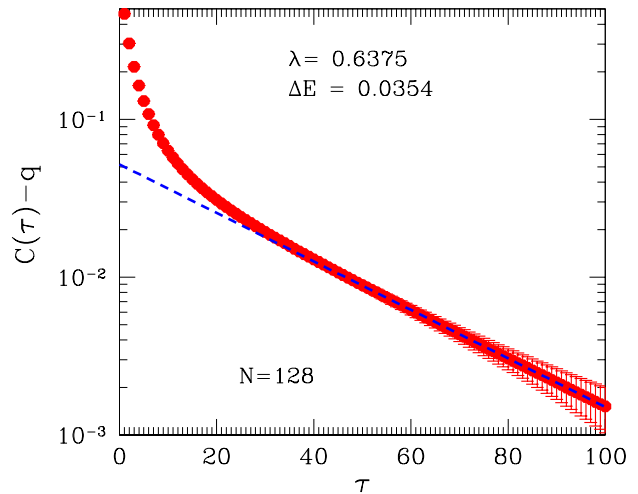}
\caption{(Color online) A log-linear plot of the time dependent
correlation function for an instance with $N=128$ near the minimum gap.
The energy gap is the negative of the slope at large values of $\tau$.
The number of time slices was $L_\tau = 300$. The error bars were estimated by
repeating the runs many (typically 100) times.
\label{fig:ctau} }
\end{center}
\end{figure}

We calculate the time-dependent
correlation function
\vspace{-0.2cm}
\begin{equation}
C(\tau) = {1 \over N' L_\tau} \sum_{i=1}^{N'} \sum_{\tau_0=1}^{L_\tau}
\langle\, \sigma^z_i(\tau_0 + \tau) \sigma^z_i(\tau_0)\, \rangle \, ,
\end{equation}
with $ \Delta \tau = 1 $ and $L_\tau$ large enough that $\beta \Delta E
\gg 1$, so the system is in the ground state.
For $\tau \ll \beta$, the correlation function $C(\tau)$ will be a sum
of exponentials
\begin{equation}
C(\tau) = q + \sum_{n\geq 1} A_n
\exp[-(E_n - E_0) \tau] \, , \label{ctau}
\end{equation}
where the $A_n$
are constants and $q$, the long time limit of the correlation function,
is \textit{determined} from
\begin{equation}
q = {1 \over N'}
\sum_{i=1}^{N'} \left( {1 \over L_\tau}\, \sum_{\tau_0=1}^{L_\tau}
\langle \, \sigma^z_i(\tau_0) \, \rangle \right)^2 \, , \label{q}
\end{equation} 
At large $\tau$, the sum in Eq.~(\ref{ctau}) is dominated
by the term corresponding to the first excited state, ($n=1$), and so
$\Delta E$ can be obtained by fitting $\log [C(\tau) - q]$ against
$\tau$ for large $\tau$. Figure \ref{fig:ctau} shows such a fit for an
instance with $N=128$ near the minimum gap.

We determine $\Delta E_{\rm min}$, the \textit{minimum} value of the gap
(to the first excited state), as $\lambda$ is varied. 
Fig.~\ref{fig:DE_1} shows QMC results for the gap between the ground
state and the first excited state as a function of the control parameter
$\lambda$ for one instance with $N=64$. The inset shows more clearly the
region of the minimum gap. The gap is greater than zero for both $\lambda = 0$ and
1 (a property of this model) and is much smaller at an intermediate
value $\lambda^*$ in the vicinity of the quantum phase transition. Each
instance has to be carefully monitored to find the minimum gap, since
$\lambda^*$ is different for each instance.

For the largest size studied, $N = 128$, we found that for some
instances it was difficult to determine $q$ accurately for a range of
$\lambda$, because the simulation was not fully equilibrated; the required
numnber of sweeps increases rapidly with $N$. As a result,
plots of $C(\tau)-q$, see Fig.~\ref{fig:ctau}, were
strongly curved.
In a few cases the error in the
computed value of $q$ was small and the problem could be cured by
allowing $q$ to vary slightly away from the computed value when doing
the fits. However, we did not trust this procedure if the correction to
$q$ was large.  For the remaining 13 out of 50 instances, we were able to provide an
upper bound for the minimum gap (from the range of $\lambda$ where $q$
was successfully computed) and this turned out to be less than our
eventual estimate for the median gap. Hence we were able to obtain
reliable data for sizes up to $N=128$. However, at present we are not
able to study much larger sizes because of the difficulty in determining
$q$. 

\begin{figure}
\begin{center}
\includegraphics[width=\columnwidth]{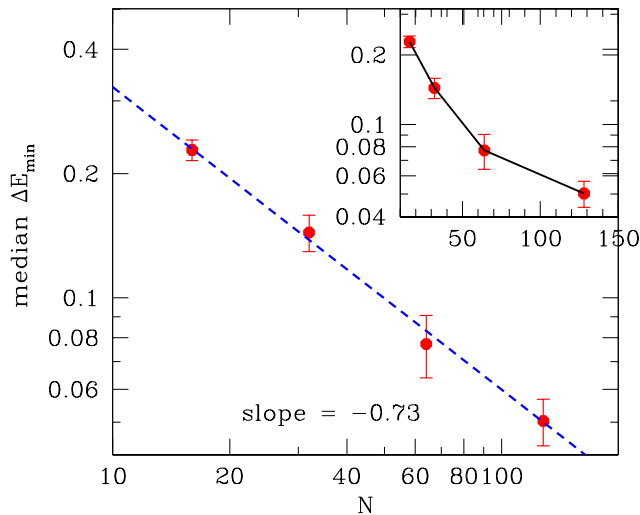}
\end{center}

\vspace{-0.5cm}
\caption{(Color online) A log-log plot of the median of the minimum gap
as a function of the number of bits $N$ up to $N=128$. From the
satisfactory straight line fit, it is seen that the median $\Delta
E_{\rm min}$ \textit{decreases as a power law}, $N^{-\mu}$ with $\mu =
0.73 \pm 0.06$. The number of instances is 50 except for $N=64$ for
which it is 45. The inset shows a log-linear plot. The pronounced
curvature shows that the behavior is \textit{not exponential} for this
range of sizes, in contrast to the classical DPLL algorithm, data for
which is shown in Fig.~\ref{fig:DPLL}. }
\vspace{-0.5cm}
\label{fig:DE_min_median}
\end{figure}

Since we are interested in the \textit{typical} minimum gap (among
different instances), rather than the average (or smallest) we show in
Fig.~\ref{fig:DE_min_median} the \textit{median} of the minimum gap for
$N \le 128$. The main figure is a log-log plot, and the dashed line
corresponds to the median $\Delta E_{\rm min}$ varying as $N^{-0.73}$.
The pronounced curvature in the inset (log-linear plot) shows that the
behavior is \textit{not} exponential. The minimum gap therefore follows
a power law for this range of sizes, implying polynomial complexity.
This result is consistent with that found by Farhi et
al.~\cite{farhi:01b} and Hogg~\cite{hogg:03} for much smaller sizes
($N\lesssim 20$--$24$).
Ba\~nuls et al.~\cite{banuls:06} studied the QAA using
using matrix product states for sizes up to $N=60$, but their result that the
complexity becomes \textit{independent} of size for $N \gtrsim 40$ is
surprising and quite different from ours.

\begin{figure}[t]
\begin{center}
\includegraphics[width=\columnwidth]{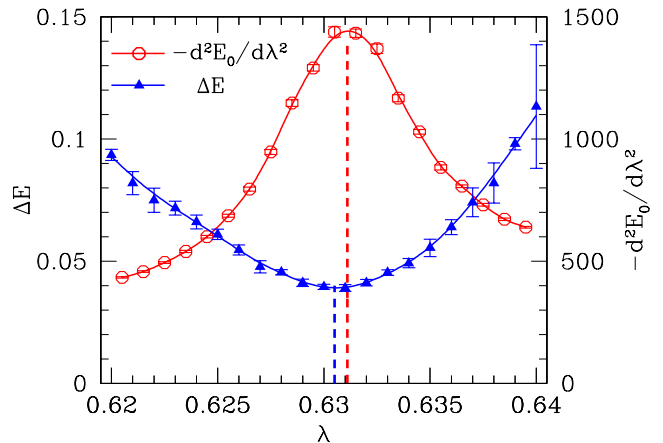}

\vspace{-0.5cm}
\caption{The gap $\Delta E(\lambda)$ (blue), and 
$-d^2 E_{0}/d\lambda^2$ (red), against $\lambda$ for an
instance with $N=128$. Solid lines are cubic interpolations. The
location of the minimum gap, $\lambda^*=0.6306$, is, within margin of
error, equal to the maximum of $-d^2 E_{0}/d\lambda^2$ at $\lambda = 0.6311$
(both shown by vertical dashed lines).
\label{fig:deriv}}

\vspace{-1cm}
\end{center}
\end{figure}

In addition to the energy gap $\Delta E(\lambda)$, we also
investigated $-d^2 E_{0}/d\lambda^2 =
2\sum_{m=1}^{2^N}|V_{0m}|^2/(E_m-E_0)$ since this gives additional
information about matrix elements
near the
avoided crossing point $\lambda^{*}$. We  determined this from
$\chi=\int_0^\beta \left\langle \left[
\mathcal{H}_{\rm P}(\tau) \mathcal{H}_{\rm P}(0) - \langle
\mathcal{H}_{\rm P} \rangle^2 \right] \right\rangle
d\tau=-(1-\lambda)^2 d^2 E_{0}/d\lambda^2$, finding that $V_{0m}$
depends on $N$ very weakly near $\lambda\simeq \lambda^*$. We also found
that the location of the maximum of $-d^2 E_{0}/d\lambda^2$ coincides to
a good precision with $\lambda^*$, see Fig.~\ref{fig:deriv}. Hence
the sum in the expression for $d^2 E_{0}/d\lambda^2$ is dominated by its
first term ($m$=1) in the vicinity of the avoided-crossing at $\lambda^*$,
which is of the Landau-Zener type (collision of $E_1$ and $E_0$ levels
only). This suggests that ${\cal T}={\hbar |V_{10}(\lambda^*)| /
[\varepsilon  \left(\Delta E_{\rm min}\right)^2]}$ is an accurate estimate
for the algorithm complexity, where $\varepsilon\ll 1$ is an
$N$-independent constant. As a result, ${\cal T}\sim N^{2\mu}$ where
$\mu=0.73\pm 0.06$.

To conclude, by using QMC simulations we have considerably extended the
range of sizes over which the complexity of the quantum adiabatic
algorithm (QAA) can be investigated. For sizes up to $N$ =128, where the
benchmark classical algorithm for satisfiability problems (DPLL) shows
exponential complexity, the QAA shows polynomial behavior of the
\textit{median} minimum gap, and hence presumably polynomial behavior of the
median complexity (contrast Fig.~\ref{fig:DE_min_median} with
Fig.~\ref{fig:DPLL}).  However, our results for the median do not rule out
the possibility that 
some instances have exponential complexity.  We also found a Landau-Zener
(pairwise) character of the avoided crossing at the minimum gap point.


\begin{acknowledgments}
The work of APY is supported by NSF Grant No.~DMR 0337049 and by a
generous allocation of computer time from the Hierarchical Systems
Research Foundation. The work of SIK and VNS is supported by US
National Security Agency's Laboratory of Physics Sciences and NASA ARC
NAS Supercomputing Center. 
\end{acknowledgments}

\bibliography{refs}

\end{document}